\documentclass[a4paper,USenglish]{lipics-v2021}
\usepackage[ruled,vlined,linesnumbered]{algorithm2e}
\usepackage{multicol,multirow,tabularx}
\usepackage{listings}
\usepackage{floatrow, subcaption, wrapfig}
\usepackage{csquotes}
\usepackage{tlatex}
\newcommand{\code}[1]{\texttt{#1}}

\newfloatcommand{capbtabbox}{table}

\author{Jacob Nelson-Slivon}{Lehigh University}{jjn217@lehigh.edu}{}{}{}
\author{Lewis Tseng}{Boston College}{lewis.tseng@bc.edu}{}{}{}
\author{Roberto Palmieri}{Lehigh University}{rop417@lehigh.edu}{}{}{}
\authorrunning{J. Nelson-Slivon, L. Tseng, R. Palmieri}
\Copyright{Jacob Nelson-Slivon, Lewis Tseng, Roberto Palmieri}
\nolinenumbers 

\begin{document}

\title{Technical Report: Asymmetric Mutual Exclusion for RDMA}
\maketitle

\begin{displayquote}
\small
``New instances of the mutual exclusion problem can still arise that make currently impractical algorithms useful, or that require new algorithms. One source of those instances is new hardware designs.'' -- Leslie Lamport~\cite{lamport2021pluscal}
% ``New instances of the mutual exclusion problem can still arise that make currently impractical algorithms useful... One source of those instances is new hardware designs.'' -- Leslie Lamport~\cite{lamport2021pluscal}
\end{displayquote}

\begin{abstract}
Coordinating concurrent access to a shared resource using mutual exclusion is a fundamental problem in computation.
In this paper, we present a novel approach to mutual exclusion designed specifically for distributed systems leveraging a popular network communication technology, remote direct memory access (RDMA).
Our approach enables local processes to avoid using RDMA operations entirely, limits the number of RDMA operations required by remote processes, and guarantees both starvation-freedom and fairness.

\end{abstract}

\vspace{-5pt}
\section{Introduction}
\label{sec:intro}

% Coordinating access to shared resources is a fundamental challenge in concurrent computation and has motivated the widespread adoption of hardware enabled synchronization mechanism like compare-and-swap (CAS), whose importance in concurrent computing is indubitable.
% An atomic CAS operation enables an arbitrary number of threads to agree on a value in a wait-free manner~\cite{todo}, making it a powerful tool for building mutual exclusion primitives and lock-free algorithms~\cite{todo}.

% The value of atomic CAS operations is likewise evident in the increasingly popular network communication technology, remote direct memory access (RDMA), which directly implements the shared memory abstraction with hardware in the distributed setting by allowing a process to access memory on a remote machine without interacting with another process~\cite{todo}.

Remote direct memory access (RDMA) is a popular network communication technology that directly implements the shared memory abstraction in the distributed setting by allowing a process to access memory on a remote machine \textit{without} interacting with another process~\cite{technologies2015rdma, 2007infiniband, gzc2016gen}.
In addition to the ability to read and write shared memory on another machine, RDMA also enables processes to perform atomic read-modify-write (RMW) operations on remote memory, like compare-and-swap (CAS) and fetch-and-add (FAA).
Hence, the API closely resembles that of modern shared-memory architectures.
% In practice, RDMA CAS is used in nearly all distributed systems leveraging RDMA to synchronize concurrent access~\cite{todo}.

\begin{wraptable}{R}{.4\textwidth}
    \small
    \centering
    \begin{tabular}{|cc|ccc|}
        \hline
        \multicolumn{2}{|c|}{\multirow{2}{*}{Access (8B)}} & \multicolumn{3}{c|}{Remote (RDMA)} \\
        \cline{3-5}
        & & \code{Read} & \code{Write} & \code{RMW}  \\
        \hline
        \multirow{3}{*}{\rotatebox[origin=c]{90}{Local}} & \multicolumn{1}{|c|}{\code{Read}} & Yes & Yes & Yes \\
        & \multicolumn{1}{|c|}{\code{Write}} & Yes & Yes & \color{red}{No} \\
        & \multicolumn{1}{|c|}{\code{RMW}} & Yes & Yes & \color{red}{No} \\
        \hline
    \end{tabular}
    \vspace{-3pt}
    \caption{Atomicity between 8-byte local and remote accesses.}
    \label{tab:atomicity}
\end{wraptable}

% \begin{figure*}
% \begin{floatrow}
% \capbtabbox{
%     \small
%     \centering
%     \begin{tabular}{|cc|ccc|}
%         \hline
%         \multicolumn{2}{|c|}{\multirow{2}{*}{Access}} & \multicolumn{3}{c|}{Remote} \\
%         \cline{3-5}
%         & & \code{Read} & \code{Write} & \code{RMW}  \\
%         \hline
%         \multirow{3}{*}{\rotatebox[origin=c]{90}{Local}} & \multicolumn{1}{|c|}{\code{Read}} & Yes & Yes & Yes \\
%         & \multicolumn{1}{|c|}{\code{Write}} & Yes & Yes & \color{red}{No} \\
%         & \multicolumn{1}{|c|}{\code{RMW}} & Yes & Yes & \color{red}{No} \\
%         \hline
%     \end{tabular}
% } {
%     \vspace{20pt}
%     \caption{Atomicity between local and remote operations in an RDMA-based distributed system.}
%     \label{tab:atomicity}
% }
% \ffigbox{
%     \includegraphics[width=0.4\textwidth]{DISC 22/figures/cas_scalability.png}
%     }{
%     \caption{The per-client (left axis) and total (right axis) throughput (ops/s) of RDMA compare-and-swap operations on a register.}
%     \label{fig:cas}
%   } 
% \end{floatrow}
% \end{figure*}

Also similar to modern architectures, the memory semantics of RDMA is \textit{not} sequentially consistent.
% equivalent to traditional shared memory.
% For instance, writes to a given memory location may be reordered in unexpected ways, leading to a memory order that differs from even the most relaxed memory ordering in modern architectures~\cite{dan2016modeling}.
Since remote operations complete asynchronously, local and remote access to a given memory location may be reordered~\cite{dan2016modeling}.
Thus, the programming model requires that programmers wait until remote operations complete and use the provided RDMA memory fences, along with local ones, to guarantee ordering\footnote{Both of these behaviors are assumed throughout the paper}~\cite{2007infiniband}.
% \todo{The last part is confusing because you move from a description to an assumption. Maybe we separate them or add a footnote?}
% \lewis{I don't quite understand this sentence . So applications cannot even ensure sequential consistency? No Read-My-Write?}

Moreover, atomicity between local and remote accesses is not guaranteed.
For instance, RDMA reads and writes are only atomic with their local counterparts for accesses within a cache line~\cite{dragojevic2014farm}.
Also, remote RMW operations are not atomic with local RMW operations in commodity hardware without global atomicity support (i.e., atomicity among all operations -- see Table~\ref{tab:atomicity}).
% Essentially, a remote RMW is nothing more than a read then write from the perspective of local memory.
For instance, many RDMA implementations enforce atomicity of the remote RMW operations in the RDMA-capable network interface controller (RNIC)~\cite{kalia2016design}, precluding global atomicity.
% not the same underlying mechanism as the local accesses (e.g., cache-line locking~\cite{intelcorporation2021intel}).
% We formalize differences between local and remote processes with the notion of \emph{operation asymmetry}, a property that captures the fact that not all processes can perform the same set of operations on an object.
An upshot of the lack of atomicity for RMW operations is that synchronizing local and remote processes becomes more difficult and costly.
In practice, RDMA RWM operations are frequently used in RDMA-based distributed systems when synchronizing concurrent access among processes (e.g., ~\cite{chen2017fast, binnig2016end, zamanian2017end, yoon2018distributed}).
In these systems, all processes rely on the RNIC to provide consistency.
More precisely, local processes must use the \emph{loopback} mechanism, which allows a process to access memory on its own machine by passing through the RNIC.
Although fast compared to other network communication, RDMA is still at least an order of magnitude slower than local accesses~\cite{nelson2020performance, kalia2016design}, degrading performance for local processes.
Additionally, the RDMA loopback mechanism can cause poor performance due to internal congestion~\cite{kong2022collie}.
Alternatively, systems can leverage remote procedure calls (RPCs) to allow all synchronization to be handled exclusively with local processes.
However, sending messages nullifies the performance benefit of directly accessing remote memory that RDMA provides.
The prevalence of RPCs in RDMA-based systems (e.g., ~\cite{dragojevic2014farm, ziegler2019designing, kalia2016fasst, mitchell2013using, wang2015hydradb}) is attributed in part to the challenges associated with synchronizing local and remote processes.
% by performing an RDMA operation on local memory, which has consequences.
% For most deployments, the standard approach is to have all processes\lewis{, including local processes,} use RDMA when performing atomic RMW operations to ensure that they are consistent with remote accesses, using RDMA loopback if necessary. \lewis{For people with no RDMA background, loopback might need some explanation?}
% As is the case with traditional networking technologies, RDMA offers a loopback mechanism that local processes may use.

% Finally, Figure~\ref{fig:cas} demonstrates how the scalability of RDMA CAS operations is limited due to resource contention, even in the absence of loopback traffic.
% Throughput cannot scale beyond two clients per node in a cluster that has 1 server and 10 client nodes, communicating via a Mellanox ConnectX-4 RNIC~\cite{2018mellanox} .

In this paper, we describe a new mutual exclusion algorithm designed to capture the nuanced requirements of synchronizing local and remote processes in RDMA-aware systems without using loopback or RPCs.
First, we define a model to precisely capture the interaction between local and remote processes and formalize the concept of \textit{operation asymmetry}, a property that captures how processes interact with registers.
This model motivates and informs the design of our RDMA-aware synchronization primitive that jointly optimizes behavior for \textit{both} local and remote processes by eliminating RDMA use in local processes and limiting the number of RDMA operations required by remote processes.
To the best of our knowledge, we are the first to solve mutual exclusion specifically for RDMA in a manner that does \textit{not} require loopback or RPC handling for local processes, is \emph{starvation-free} (i.e., a calling process eventually executes its critical section~\cite{herlihy2021}), and is \emph{fair} (i.e., lock acquisitions are first-come-first-served~\cite{herlihy2021}).

To achieve our goal, we draw from lock cohorting~\cite{dice2012lock} and the classic Peterson's lock algorithm~\cite{peterson1981myths} to guarantee mutual exclusion between a local process $p_L$ and remote process $p_R$.
The processes $p_L$ and $p_R$ are elected by the set of local processes and remote processes, respectively, via orthogonal mechanisms that we embed directly into Peterson's lock.
Thus, we reduce the $n$-process mutual exclusion problem to synchronizing a single local process and a single remote process.
Compared to our approach, previous strategies to synchronize local and remote processes without RPCs are \textit{not} starvation-free, entail further hardware support, or require RDMA loopback, which we discuss later.
Finally, to verify the correctness of our solution we model check a TLA+ specification translated from a PlusCal algorithm.
% \lewis{Do prior approaches use loopback or need to handle RPC? If yes, it might be useful to emphasize it again here. Becauase at this point, it's not easy to see what `generalizable' means. Also, it's good to mention the notion of fair here, but personally, I'm not a big fan of throwing out new idea/term at the very last paragraph.}
% In contrast, our approach is both starvation-free and applicable to any RDMA-enabled system.

\vspace{-5pt}
\section{System Model}

We model an RDMA-based distributed system as a set of nodes $N$ and processes $P$, for which $p_i^j$ is a process running on node $n_i$ with a process identifier $j$.
All processes can access an RDMA-accessible shared memory $M$ partitioned among the nodes, where each partition $m_i$ on node $n_i$ is composed of atomic registers.
A register $r_i^j$ resides in memory partition $m_i$ and is identified by $j$.
% \st{Furthermore,} a register may be accessed in one of two ways that is defined by the underlying mechanism by which the access occurs, forming the basis for our definition of asymmetry.\todo{let's try to reword this last sentence together}
Similar to previous RDMA system models~\cite{aguilera2018passing}, we assume that processes are asynchronous and that accesses to memory are failure free.

% The notion of locality between a process and register is critical to our model, which motivates distinguishing the placement of processes and registers using the concept of nodes.
We define locality as a relation between the set of processes and the set of registers in our system.
A process is \emph{local} if and only if it resides on the same node as the given register.
Otherwise, it is \emph{remote}.
A \emph{local access} is the means by which a local process accesses a register via the traditional memory subsystem of a machine (i.e., no RDMA operation).
Alternatively, a \emph{remote access} is an operation on a register that passes through the RNIC.
For a given process, an operation on a register is \emph{enabled} if the process is able to access the register using the given operation.
Intuitively, local accesses are only enabled for local processes.
However, remote accesses are enabled for all processes because RDMA allows processes to access a local register using RDMA loopback.

For each class of access (local or remote), a register supports three operations: read, write and compare-and-swap.
We denote the local operations using \code{Read}, \code{Write} and \code{CAS}, and the remote operations with \code{rRead}, \code{rWrite} and \code{rCAS}.
Recall from Table~\ref{tab:atomicity}, the atomicity of operations between classes is not guaranteed.
Due to this behavior, an \code{rCAS} operation appears to a local process as if it were a \code{Read} then \code{Write}. Without loss of generality, we define an object as \emph{operation asymmetric} if, given two processes, the intersection of their respective enabled operations on the object is \emph{not equal} to their union.
In our model, registers are operation asymmetric since remote processes are constrained to remote accesses.
% Similarly, if we model a remote process as a local process with only \code{Read} and \code{Write} enabled the register is still operation asymmetric since \code{CAS} is not enabled for the modeled process. \lewis{Not sure if we want/need the final sentence?}

\vspace{-5pt}
\section{Mutual Exclusion Under Operation Asymmetry}

Mutual exclusion is a known problem in which access to a shared resource is coordinated among two or more concurrent processes~\cite{lynch1996distributed}.
% Specifically, at most one process may execute its critical section at a time.
% CAS-based mutual exclusion primitives harness the power of atomic CAS operations, which stems from the fact that any number of processes can decide on a value using a single compare-and-swap register.
A naive solution to mutual exclusion is to implement a lock by enforcing that all processes, including the local ones, utilize \code{rCAS} to guarantee atomicity.
As discussed in Section~\ref{sec:intro}, RDMA loopback adds overhead and may exhibit anomalous performance bugs, causing the RNIC to become a bottleneck.
To avoid these performance pitfalls, in our solution we enforce that local processes \emph{only use local operations}.
However, since atomicity is not guaranteed among local and remote RMW operations we must reassess how to implement a mutual exclusion primitive.
Due to operation asymmetry, we require an algorithm solely built from the greatest common denominator: read-write registers.
A natural choice is to look to Peterson's lock for inspiration.

Peterson's lock~\cite{peterson1981myths} is a well-known starvation-free mutual exclusion primitive for two processes.
Because remotely accessible memory is composed of atomic read-write registers, Peterson's lock can be implemented directly over RDMA, with the appropriate memory fences, to coordinate access between a local and a remote process.
To allow multi-process synchronization, we modify the original Peterson's lock algorithm to embed an orthogonal mutual exclusion primitive whereby a process participates in the Peterson's lock protocol by first obtaining the embedded lock.
Since we wish to limit the number of remote operations required for remote processes (e.g., spinning on remote memory), we embed the widely used MCS queue lock~\cite{mellorcrummey1991algorithms}, allowing processes to spin locally while waiting in a queue of processes to acquire the lock.
This combination of locks has similar characteristics to lock cohorting~\cite{dice2012lock}, which decouples the synchronization of cohorts of processes with global synchronization.
The locks used for each step are called the \emph{cohort} and \emph{global} locks respectively, a nomenclature that we adopt.
In our approach, processes of the same locality (local or remote) compete amongst themselves using a MCS queue cohort lock to determine a leader that then participates in the global Peterson's lock protocol.

Note that a naive solution to multi-processes mutual exclusion is a \emph{filter lock}~\cite{peterson1981myths}, which extends Peterson's lock for multiple processes.
Briefly, processes compete for access to successive levels that each hold back one process.
The number of levels is equal to one less than the number of processes that might acquire the lock.
Unfortunately, this would require both remote spinning and a number of remote accesses proportional to the number of processes that might contend for the lock, even if a process executes in isolation.
Lamport's Bakery algorithm~\cite{herlihy2021} also demonstrates the same undesirable behavior for remote processes. 

Below, we first discuss our modified Peterson's lock mechanism, then describe our MCS queue lock design.
Subsequently, we address fairness by extending the Peterson's lock API and incorporating a budget into the cohort lock.
Finally, we compare our technique to lock cohorting and other RDMA-based mutual exclusion approaches. 
Algorithm~\ref{algo:cohorted} and Algorithm~\ref{algo:mcs} provide the pseudo-code for our modified Peterson's lock and RDMA-based MCS queue lock implementations, respectively.
Appendix~\ref{app:tla} contains a PlusCal version of our design that can be translated to a TLA+ specification, which we model checked.

\begin{algorithm}
\scriptsize
\caption{Modified Peterson's Lock}
\label{algo:cohorted}
\DontPrintSemicolon
\KwData{(global) \code{cohort[2]}, \code{victim}}
\texttt{pLock()}
\Begin{
    \code{id} $\gets$ \code{getCid()} \tcp*{Get ID of process class}
    \code{other} $\gets 1 -$ \code{id} \\
    \code{isLeader} $\gets$ \code{cohort[id].qLock()} \\
    \If{\upshape \code{isLeader}}{
        \code{victim} $\gets$ \code{id} \\
        \While{\upshape \code{cohort[other].qIsLocked()} \textbf{and} \code{victim} $ = $ \code{id}}{
        \label{line:petersons:wait}
        \textbf{wait}} 
    }
}

\texttt{pUnlock()}
\Begin{
    \code{id} $\gets$ \code{getCid()} \\
    \code{cohort[id].qUnlock()}
}

\texttt{pReacquire()}
\label{line:reaquire}
\Begin{
    \code{id} $\gets$ \code{getCid()} \\
    \code{other} $\gets 1 -$ \code{id} \\
    \code{victim} $ \gets$ \code{id} \tcp*{Yield lock to waiting process}
    \lWhile{\upshape \code{cohort[other].qIsLocked()} \textbf{and} \code{victim} $ = $ \code{id}}{\textbf{wait}\tcp*[f]{Reaquire lock}} 
}
\end{algorithm}

\subsection{Algorithm Description}
Our modified Peterson's lock algorithm (Algorithm~\ref{algo:cohorted}) has two global variables: \code{cohort}, which is a two element array of cohort locks, and \code{victim}, which decides which process yields execution.
In \code{pLock()}, a process first announces interest in executing its critical section by locking the appropriate cohort lock.
When the cohort lock acquisition returns \code{true}, the calling process directly acquired the lock from another member of its cohort and may enter the critical section without additional steps.
Otherwise, the process must engage in the Peterson's lock protocol, in which case the wait condition (Line~\ref{line:petersons:wait}) is nearly identical to the original algorithm other than the calling process checking whether the other cohort lock is held.
\code{pUnlock()} simply unlocks the cohort lock corresponding to the calling process's class, allowing a waiting process to proceed.
Crucially, any interleaving of instructions in concurrent calls to \code{pLock()} will only allow a single process access to the critical section, assuming that sequential consistency is enforced.

\begin{algorithm}
\scriptsize
\caption{Budgeted MCS Queue Lock for Remote Process}
\label{algo:mcs}
\DontPrintSemicolon
\KwData{(constants) \code{kInitBudget}; (global) \code{glock}, \code{tail}; (process-local) \code{desc}, \code{leader}}
\code{qLock()}
\Begin{
    \code{desc}  $\gets$ \code{MCSDescriptor\{kInitBudget,} $nullptr$\code{\}} \tcp*{Remotely accessible}
    \code{curr} $\gets nullptr$ \\
    \While{true}{
        \label{line:swap}
        \tcp{Note: $curr$ updated on \code{rCAS}}
        \If{\upshape \code{rCAS(tail, /*expected=*/curr, /*swap=*/\&desc)} $=$ \code{curr}}{
            \lIf{\upshape \code{curr} $=nullptr$}{\Return $true$}
             \textbf{break}
        }
    }
    \label{line:leader}
    \code{desc.budget} $\gets -1$ \\
    \code{rWrite(\&(curr->next), \&desc)} \\
    \lWhile{\upshape \code{desc.budget} $=-1$}{\textbf{wait}
    \tcp*[f]{Busy wait locally}}
    \label{line:wait}
    \If{\upshape \code{desc.budget} $=0$}{\code{glock.pReacquire()} \tcp*{Provides fairness}
    \code{desc.budget} $\gets$ \code{kInitBudget}}
    
    \Return $false$
}

\texttt{qUnlock()}
\Begin{
    \If{\upshape \code{desc.next} $=nullptr$}{
        \lIf{\upshape \code{rCAS(tail, \&desc, $nullptr$)} $=$ \code{\&desc}}{
            \Return
        }
        \lWhile{\upshape \code{desc.next} $=nullptr$}{\textbf{wait}}
    }
    \code{rWrite(\&(desc.next->budget), desc.budget - 1)}\tcp*[f]{Pass the lock}
}

\texttt{qIsLocked()}
\Begin{
    \Return \code{rRead(tail)} $\ne nullptr$
}
\end{algorithm}

Next, we describe our modified MCS queue algorithm (Algorithm~\ref{algo:mcs}) for remote processes.
It should be noted that the local version of the algorithm can be obtained by directly replacing each remote access with a local one.
This algorithm maintains two global variables: \code{glock} is a remote reference to the global lock, \code{tail} is a remote reference to the corresponding slot in the \code{cohort} array of Algorithm~\ref{algo:cohorted}, which acts as the tail of the lock queue.
In \code{qLock()}, a process atomically swaps a new descriptor into \code{tail} then waits until the process is at the head of the queue, finally returning whether the queue was empty at its outset.
The value swapped into the tail of the queue contains an address of the remotely accessible descriptor \code{desc}, which is local to the locking process.
If \code{tail} was not previously set, the call returns \code{false} since the cohort lock was not passed to the calling process.
% with the while loop at Line~\ref{line:swap}.
Otherwise, \code{desc} allows the locking process to spin locally while waiting for another process to pass the lock via an \code{rWrite} operation, returning \code{true} once the lock is passed.
When the queue is empty, a lone process requires only a single \code{rCAS} to acquire the lock.
Otherwise, if the queue is not empty, a lone process requires an additional \code{rWrite} to set its predecessor's next value.
Note that once the descriptor is enqueued the calling process avoids remote spinning, thus reducing network traffic.

After acquiring the lock and performing its critical section, a process attempts to release the lock following the conventional MCS queue algorithm, which tries to CAS the back of the queue back to a null value.
Recall that if the CAS operation in the MCS queue lock \code{qUnlock()} is successful, then it also releases the Peterson's lock, since the corresponding \code{cohort} is now unset.
Otherwise, the process passes the lock to the next waiting process by performing a \code{rWrite} to the location returned by the attempted CAS.
At worst, a process requires an \code{rCAS} operation followed by an \code{rWrite} when unlocking.

The above algorithm is unfair because the lock may be passed indefinitely among processes of the same class.
To address this, we extend the Peterson's lock to support \code{pReacquire()} (Algorithm~\ref{algo:cohorted} Line~\ref{line:reaquire}), which releases the lock by setting itself as the \code{victim} then immediately reacquires the lock.
We also alter the original MCS queue algorithm to support a budget, similar to the technique used by Dice et al.~\cite{dice2012lock}.
A lock is passed by setting the budget of a waiting process to a non-negative integer that represents the number of remaining lock acquisitions.
When the budget reaches zero, an acquiring process must call \code{pReacquire()} on the global lock.
If there is a waiting process of the opposite class, it will be allowed to proceed.
Otherwise, the calling process reacquires the global lock and resets the budget.
Since the global lock is released after a bounded number of cohort lock acquisitions, and the global lock is itself fair (i.e., a waiting process cannot be overtaken), our approach is fair~\cite{dice2012lock}.

In summary, our technique enables local and remote processes to synchronize via independent mutual exclusion primitives embedded in a modified Peterson's lock algorithm.
By using a budgeted MCS queue lock as the embedded lock, we can provide fairness.
Lastly, our approach is RDMA-aware since local processes avoid RDMA loopback and remote processes avoid remote spinning, both important factors in RDMA-enabled system performance.

% \lewis{Looks good to me overall. If we have space, it might be a good idea to add one more paragraph to summarize why it is RDMA-friendly.This will help the reviewers identify key techniques/innovations.}

\vspace{-5pt}
\section{Related Work}
\label{sec:discussion}

Similar to lock cohorting~\cite{dice2012lock}, a strategy for NUMA-aware synchronization, our approach allows a group of processes to compete amongst itself before acquiring a global lock.
Our technique explicitly couples the global and cohort locks to achieve behavior tailored to the respective processes in our system.
By embedding the cohort lock in the global lock, we avoid an additional remote access for remote processes while maintaining the integrity of the global lock.
The application of lock cohorting in a distributed setting is a natural extension of the technique but it requires rethinking the design to optimize for operation asymmetry between local and remote processes, yielding a lock primitive that is of independent interest.

To the best of our knowledge, our approach is the first mutual exclusion primitive designed for RMDA that provides local-only access for local processes while maintaining fairness and avoiding RPCs.
A notable alternative is the technique pioneered by Wei et al.~\cite{wei2020fast}, which allows local accesses to be protected by hardware transactional memory (HTM) while remote accesses acquire a lock using RDMA CAS.
This technique only applies to architectures supporting HTM, which is increasingly disabled due to security concerns~\cite{lipp2018meltdown, intel2021performance}.
Due to cache coherent I/O, a local hardware transaction is aborted whenever a remote process acquires the lock.
Local operations use local accesses in the common case but a fallback path using RDMA is also needed.
Another potential option is to leverage RDMA-accessible memory permissioning, which atomically revokes remote access~\cite{aguilera2019impact,2007infiniband}, to devise a mutual exclusion algorithm.
However, this approach is known to be slow~\cite{aguilera2020microsecond} and is not easily made starvation-free since remote access may be continuously revoked by local processes.

% \todo{I would call this section Related Work. Discussion might be misleading. Also, I would start with lock cohorting and HTM 3rd and4th para, and then talk about AMP and diseggrated memory. I'm still debating if it's appropriate for a BA talking about that since those are performance considerations and we didn't implement our lock yet.}
Finally, our notion of operation asymmetry is a useful property to capture the relative capabilities of processes.
While in our model asymmetry stems from the operations with which a process accesses a register, this notion can also capture the behavior of \emph{asymmetric multiprocessing} (AMP).
AMP is a recent architectural innovation in which cores on the same processor offer different power characteristics~\cite{liu2022asymmetry}.
So called \emph{large cores} specialize in higher performance while \emph{small cores} are energy efficient.
The adoption of AMP in commodity hardware has led to new techniques to balance their respective needs~\cite{liu2022asymmetry, 2019cfs}.
We leave a more detailed analysis of the application of operation asymmetry to other domains for future work but we believe it is a practical tool to capture asymmetry in process capabilities, especially as it pertains to recent trends in heterogenous architectures.

\vspace{-5pt}
\section{Conclusion}
In this paper we present a new model for describing the atomicity of memory for RDMA-based systems and we define operation asymmetry to capture how processes do not operate on a register equally.
Based on our model, we propose a fair mutual exclusion mechanism inspired by lock cohorting to enable local and remote processes to synchronize while optimizing for their individual behavioral constraints.
Our design embeds an RDMA-aware MCS queue lock into a modified version of Peterson's lock protocol to achieve our design goals.
To the best of our knowledge, our technique is the first mutual exclusion solution that allows synchronizing local and remote processes while avoiding both RDMA loopback and RPCs.

\newpage
\appendix
\section*{Appendix}
\section{PlusCal Algorithm}
\label{app:tla}
\definecolor{boxshade}{gray}{0.85}

\tlatex
\small
\setboolean{shading}{true}
\@x{}\moduleLeftDash\@xx{ {\MODULE} qplock}\moduleRightDash\@xx{}%
\@x{ {\EXTENDS} Integers ,\, Sequences ,\, TLC}%
\@x{ {\CONSTANTS} NumProcesses ,\, InitialBudget}%
\@x{ {\ASSUME} NumProcesses \.{>} 0}%
\@x{ {\ASSUME} InitialBudget \.{>} 0}%
\@x{ NP \.{\defeq} NumProcesses}%
\@x{ B \.{\defeq} InitialBudget}%
\@pvspace{8.0pt}%
\pcalsymbolstrue
\csyntaxfalse
\@x{\@s{8.06} {\p@mmalgorithm} qplock}%
\@x{ {\p@variables}}%
\@x{\@s{16.4}}%
\@y{\@s{0}%
 Global
}%
\@xx{}%
\@x{\@s{16.4} victim\@s{0.30} \.{\in} \{ 1 ,\, 2 \} ,\,}%
\@x{\@s{16.4} cohort \.{=} [ x \.{\in} \{ 1 ,\, 2 \} \.{\mapsto} 0 ] ,\,}%
 \@x{\@s{16.4} descriptor \.{=} [ x \.{\in} ProcSet \.{\mapsto} [ budget
 \.{\mapsto} \.{-} 1 ,\, next \.{\mapsto} 0 ] ] ,\,}%
\@x{\@s{16.4}}%
\@y{\@s{0}%
 Process-local
}%
\@xx{}%
\@x{\@s{16.4} passed \.{=} [ x \.{\in} ProcSet \.{\mapsto} {\FALSE} ] ,\,}%
\@pvspace{8.0pt}%
\@x{ {\p@define}}%
\@x{\@s{16.4} Us ( pid ) \.{\defeq} ( pid \.{\%} 2 ) \.{+} 1}%
\@x{\@s{16.4} Them ( pid ) \.{\defeq} ( ( pid \.{+} 1 ) \.{\%} 2 ) \.{+} 1}%
\@x{\@s{16.4} Budget ( pid ) \.{\defeq} descriptor [ pid ] . budget}%
\@x{ {\p@end} {\p@define} {\p@semicolon}}%
\@pvspace{8.0pt}%
\@x{ {\p@procedure} AcquireGlobal ( )}%
\@x{ {\p@begin}}%
\@x{\@s{16.4} g1\@s{.5}\textrm{:}\@s{3} victim \.{:=} self {\p@semicolon}}%
\@x{\@s{16.4} gwait\@s{.5}\textrm{:}\@s{3} {\p@while} {\TRUE} {\p@do}}%
 \@x{\@s{50.47} g2\@s{.5}\textrm{:}\@s{3} {\p@if} cohort [ Them ( self ) ]
 \.{=} 0 {\p@then}}%
\@x{\@s{81.87} {\p@goto} g4 {\p@semicolon}}%
\@x{\@s{67.23} {\p@end} {\p@if} {\p@semicolon}}%
 \@x{\@s{50.47} g3\@s{.5}\textrm{:}\@s{3} {\p@if} victim \.{\neq} self
 {\p@then}}%
\@x{\@s{81.87} {\p@goto} g4 {\p@semicolon}}%
\@x{\@s{67.23} {\p@end} {\p@if} {\p@semicolon}}%
\@x{\@s{33.16} {\p@end} {\p@while} {\p@semicolon}}%
\@x{\@s{16.4} g4\@s{.5}\textrm{:}\@s{3} {\p@return}}%
\@x{ {\p@end} {\p@procedure} {\p@semicolon}}%
\@pvspace{8.0pt}%
\@x{ {\p@procedure} AcquireCohort ( )}%
\@x{ {\p@variables} pred}%
\@x{ {\p@begin}}%
 \@x{\@s{16.4} c1\@s{.5}\textrm{:}\@s{3} descriptor [ self ] \.{:=} [ budget
 \.{\mapsto} \.{-} 1 ,\, next \.{\mapsto} 0 ] {\p@semicolon}}%
 \@x{\@s{16.4} swap\@s{.5}\textrm{:}\@s{3} pred \.{:=} cohort [ Us ( self ) ]
 {\p@semicolon} cohort [ Us ( self ) ] \.{:=} self {\p@semicolon}}%
 \@x{\@s{16.4} cwait\@s{.5}\textrm{:}\@s{3}\@s{1.99} {\p@if} {\lnot} ( pred
 \.{=} 0 ) {\p@then}}%
 \@x{\@s{32.8} c2\@s{.5}\textrm{:}\@s{3}\@s{3.22} descriptor [ pred ] . next
 \.{:=} self {\p@semicolon}}%
 \@x{\@s{32.8} c3\@s{.5}\textrm{:}\@s{3}\@s{3.22} {\p@await} Budget ( self )
 \.{\geq} 0 {\p@semicolon}}%
 \@x{\@s{32.8} c4\@s{.5}\textrm{:}\@s{3}\@s{3.22} {\p@if} Budget ( self )
 \.{=} 0 {\p@then}}%
 \@x{\@s{67.10} c5\@s{.5}\textrm{:}\@s{3} {\p@call} AcquireGlobal ( )
 {\p@semicolon}}%
 \@x{\@s{67.10} c6\@s{.5}\textrm{:}\@s{3} descriptor [ self ] . budget \.{:=}
 B {\p@semicolon}}%
\@x{\@s{52.46} {\p@end} {\p@if} {\p@semicolon}}%
 \@x{\@s{32.8} c7\@s{.5}\textrm{:}\@s{3}\@s{3.22} passed [ self ] \.{:=}
 {\TRUE} {\p@semicolon}}%
\@x{\@s{16.4} {\p@else}}%
 \@x{\@s{32.8} c8\@s{.5}\textrm{:}\@s{3} descriptor [ self ] . budget \.{:=} B
 {\p@semicolon}}%
 \@x{\@s{32.8} c9\@s{.5}\textrm{:}\@s{3} passed [ self ] \.{:=} {\FALSE}
 {\p@semicolon}}%
\@x{\@s{16.4} {\p@end} {\p@if} {\p@semicolon}}%
\@x{\@s{16.4} c10\@s{.5}\textrm{:}\@s{3}\@s{4.53} {\p@return} {\p@semicolon}}%
\@x{ {\p@end} {\p@procedure} {\p@semicolon}}%
\@pvspace{8.0pt}%
\@x{ {\p@procedure} ReleaseCohort ( )}%
\@x{ {\p@variables} size ,\, next}%
\@x{ {\p@begin}}%
 \@x{\@s{16.4} cas\@s{.5}\textrm{:}\@s{3} {\p@if} cohort [ Us ( self ) ] \.{=}
 self {\p@then}}%
\@x{\@s{32.8} cohort [ Us ( self ) ] \.{:=} 0 {\p@semicolon}}%
\@x{\@s{16.4} {\p@else}}%
 \@x{\@s{32.8} r1\@s{.5}\textrm{:}\@s{3} {\p@await} {\lnot} ( descriptor [
 self ] . next \.{=} 0 ) {\p@semicolon}}%
 \@x{\@s{32.8} r2\@s{.5}\textrm{:}\@s{3} descriptor [ descriptor [ self ] .
 next ] . budget \.{:=} Budget ( self ) \.{-} 1 {\p@semicolon}}%
\@x{\@s{16.4} {\p@end} {\p@if} {\p@semicolon}}%
\@x{\@s{16.4} r3\@s{.5}\textrm{:}\@s{3}\@s{5.31} {\p@return} {\p@semicolon}}%
\@x{ {\p@end} {\p@procedure} {\p@semicolon}}%
\@pvspace{8.0pt}%
\@x{ {\p@fair} {\p@process} p \.{\in} 1 \.{\dotdot} NP}%
\@x{ {\p@begin}}%
\@x{\@s{16.4} p1\@s{.5}\textrm{:}\@s{3} {\p@while} {\TRUE} {\p@do}}%
\@x{\@s{33.42}}%
\@y{\@s{0}%
 Non-critical section
}%
\@xx{}%
\@x{\@s{33.42} ncs\@s{.5}\textrm{:-}\@s{3} {\p@skip} {\p@semicolon}}%
\@pvspace{8.0pt}%
\@x{\@s{33.42}}%
\@y{\@s{0}%
 Acquire the cohort lock
}%
\@xx{}%
 \@x{\@s{33.42} enter\@s{.5}\textrm{:}\@s{3} {\p@call} AcquireCohort ( )
 {\p@semicolon}}%
\@pvspace{8.0pt}%
\@x{\@s{33.42}}%
\@y{\@s{0}%
 Acquire the global lock, maybe
}%
\@xx{}%
 \@x{\@s{33.42} p2\@s{.5}\textrm{:}\@s{3} {\p@if} {\lnot} passed [ self ]
 {\p@then}}%
\@x{\@s{67.64} {\p@call} AcquireGlobal ( ) {\p@semicolon}}%
\@x{\@s{50.44} {\p@end} {\p@if} {\p@semicolon}}%
\@pvspace{8.0pt}%
\@x{\@s{33.42}}%
\@y{\@s{0}%
 Critical section
}%
\@xx{}%
\@x{\@s{33.42} cs\@s{.5}\textrm{:}\@s{3} {\p@skip} {\p@semicolon}}%
\@pvspace{8.0pt}%
\@x{\@s{33.42}}%
\@y{\@s{0}%
 Release the cohort lock
}%
\@xx{}%
 \@x{\@s{33.42} exit\@s{.5}\textrm{:}\@s{3} {\p@call} ReleaseCohort ( )
 {\p@semicolon}}%
\@x{\@s{16.4} {\p@end} {\p@while} {\p@semicolon}}%
\@x{ {\p@end} {\p@process} {\p@semicolon}}%
\@x{ {\p@end} {\p@algorithm}}%
\@y{%
 ;
}%
\@xx{}%
\pcalshadingfalse \pcalsymbolsfalse
\@pvspace{8.0pt}%
\@x{}%
\@y{\@s{0}%
 Safety
}%
\@xx{}%
 \@x{ MutualExclusion \.{\defeq} ( \A\, i ,\, k \.{\in} ProcSet \.{:} ( i
 \.{\neq} k ) \.{\implies} {\lnot} ( pc [ i ] \.{=}\@w{cs} \.{\land} pc [ k ]
 \.{=}\@w{cs} ) )}%
\@pvspace{8.0pt}%
\@x{}%
\@y{\@s{0}%
 Liveness
}%
\@xx{}%
 \@x{ ExecsCriticalSectionInfinitelyOften \.{\defeq} \A\, i \.{\in} ProcSet
 \.{:} {\Box} {\Diamond} ( pc [ i ] \.{=}\@w{cs} )}%
 \@x{ StarvationFree \.{\defeq} \A\, i \.{\in} ProcSet \.{:} ( pc [ i
 ]\@s{9.22} \.{=}\@w{enter} ) \.{\leadsto} ( pc [ i ] \.{=}\@w{cs} )}%
 \@x{ DeadAndLivelockFree \.{\defeq} ( \E\, i \.{\in} ProcSet \.{:} pc [ i ]
 \.{=}\@w{enter} ) \.{\leadsto} ( \E\, i \.{\in} ProcSet \.{:} pc [ i ]
 \.{=}\@w{cs} )}%
\@pvspace{8.0pt}%
\@x{}%
\@y{\@s{0}%
 Fairness
}%
\@xx{}%
 \@x{ CohortFairness \.{\defeq} \A\, i ,\, j \.{\in} ProcSet \.{:} ( pc [ i ]
 \.{=}\@w{cwait}\@s{0.55} \.{\land} pc [ j ] \.{=}\@w{enter} ) \.{\implies} (
 pc [ i ] \.{=}\@w{cs} \.{\leadsto} pc [ j ] \.{=}\@w{cs} )}%
 \@x{ GlobalFairness\@s{2.86} \.{\defeq} \A\, i ,\, j \.{\in} ProcSet \.{:} (
 pc [ i ] \.{=}\@w{gwait} \.{\land} pc [ j ] \.{=}\@w{enter} ) \.{\implies} (
 pc [ i ] \.{=}\@w{cs} \.{\leadsto} pc [ j ] \.{=}\@w{cs} )}%
\@x{}\bottombar\@xx{}%
\endtla

\newpage
\bibliographystyle{plainurl}
\bibliography{references}

\end{document}